**Quantifying the Impact of Making and Breaking Interface Habits**

Diego Garaialde*, Christopher P. Bowers**, Charlie Pinder***, Priyal Shah*, Shashwat Parashar*, Leigh Clark*, Benjamin R. Cowan*

*School of Information & Communication Studies, University College Dublin

**Department of Computing, University of Worcester

***School of Computer Science, University of Birmingham

Abstract

The frequency with which people interact with technology means that users may develop interface habits, i.e. fast, automatic responses to stable interface cues. Design guidelines often assume that interface habits are beneficial. However, we lack quantitative evidence of how the development of habits actually affect user performance and an understanding of how changes in the interface design may affect habit development. Our work quantifies the effect of habit formation and disruption on user performance in interaction. Through a forced choice lab study task (n=19) and in the wild deployment (n=18) of a notification dialog experiment on smartphones, we show that people become more accurate and faster at option selection as they develop an interface habit. Crucially this performance gain is entirely eliminated once the habit is disrupted. We discuss reasons for this performance shift and analyse some disadvantages of interface habits, outlining general design patterns on how to both support and disrupt them.

**Quantifying the Impact of Making and Breaking Interface Habits**

**1. Introduction**

  We interact with a wide variety of devices and interfaces on a daily basis. The frequency of these interactions means that users are likely to develop habits – fast, automatic behaviours that emerge in stable contexts – around these interfaces. Although habit research has become increasing popular in fields such as HCI (Cowan, Bowers, Beale, & Pinder, 2013; Pinder, Vermeulen, Cowan, & Beale, 2018; Pinder, Vermeulen, Wicaksono, Beale, & Hendley, 2016; Renfree, Harrison, Marshall, Stawarz, & Cox, 2016; Stawarz, Cox, & Blandford, 2015), there has been less focus on how habits are formed within interfaces and how their development affects user performance. Research has instead concentrated on behaviour change interventions to impact health, work related habits (Gardner, de Bruijn, & Lally, 2011; Stawarz et al., 2015; Stawarz, Rodriguez, Cox, & Blandford, 2016), or on problematic habitual use of technology (Elhai, Levine, Dvorak, & Hall, 2017; van Deursen, Bolle, Hegner, & Kommers, 2015). Common design guidelines such as consistency (Shneiderman & Plaisant, 2010), or usability concepts such as learnability (Nielsen, 1993), allude to desirable interface properties that are beneficial for habits to develop, yet do not attempt to quantify how habit development benefits user performance. Even with optimal intentions to follow these guidelines, changes are regularly made to interfaces that incorporate new features and design norms. When this occurs, previously learned habits may become disrupted, meaning users have to relearn how to perform desired actions. Currently there is limited quantitative evidence specifically addressing how these disruptions affect performance.

  The contribution of the current paper comes from providing quantitative evidence of how the process of forming and disrupting habits affects user performance in a forced choice interaction task, similar to those seen in notification dialogs or alert boxes. To do this we conducted two studies (one in the lab and one in the wild) examining the building of interface habits. Our studies indicate that as people develop a habit, they become quicker and more accurate at option selection. A concentrated set of repetitions over 80 trials in the lab was enough to create strong improvements in speed and accuracy. In addition, more dispersed daily interactions within a 22-day period in the wild were also enough to see significant improvements in speed. Yet once these habits are disrupted, all performance benefits are significantly reduced. Our work contributes to the field by demonstrating both the benefits of habit formation and the significant cost of disrupting these habits when an interface is changed. We show that simple interface habits, like option selection, may be easily developed. When designing interfaces, designers should be cautious about interface changes, incremental or otherwise, unless clear consideration is given to the potential disruption to users' habits.

## 2. Related Work

### 2.1 What Are Habits

Around 43% of our daily activities frequently occur in consistent contexts which are conducive to habit formation (Wood, Quinn, & Kashy, 2002). Although habits are commonly defined in the literature as "learned sequences of acts that have become automatic responses to specific cues" (Verplanken & Aarts, 1999, p 104), there is currently a debate in the literature as to what truly constitutes a habit. In an effort to reduce the conflation between habits as behaviours and habits as causes of behaviours, Gardner (2015a) defines a habit as "a process by which a stimulus automatically generates an impulse towards action, based on learned stimulus-response associations" (p. 4), which isolates the description to that of impulses to act, regardless of whether an act is completed. Critically these actions and/or impulses have a degree of automaticity, and tend to be instigated during stable contexts (Orbell & Verplanken, 2010). Automaticity means that these impulses lead behaviours to be performed without intention or using limited conscious awareness (Aarts & Dijksterhuis, 2000). These become automatic when context-response associations are created in procedural memory, bypassing the need to analyse the context when searching for an appropriate response (Wood & Neal, 2007). This benefits users by decreasing the cognitive load required to complete frequently repeated actions (Law, Wehrt, Sonnentag, & Weyers, 2017) leaving more cognitive resources for other activities. While many actions are initially performed with goal-driven intentions, over time repeated exposure to stable contextual cues reinforces the association within procedural memory (Wood & Neal, 2007). This diminishes the need for the initial goal-driven motivation (Wood & Rünger, 2016) making a habit primarily driven by environmental context rather than a specific goal. As highlighted by Gardner (2015a) the process of learning stimulus-response associations is critical in forming a habit, with the strength of the stimulus-response associations dictating the strength of the developed habit.

**2.2 Interface Habits**

Interacting with technology often involves similar actions through a familiar interface in a consistent environment. Frequent interaction with these interfaces makes it likely that habits develop within this context. Recent work shows that habits form a significant part of our interaction with mobile devices. A longitudinal study on mobile device use found that users repeatedly perform checking habits, short interactions where users check on one or two applications, dispersed evenly throughout the day (Oulasvirta, Rattenbury, Ma, & Raita, 2012). The way these form is similar to the cue-behaviour-reward process seen in developing other habits (Wood & Rünger, 2016). Checking behaviours become tightly associated to particular contextual triggers, with behavioural execution leading to positive rewards such as desired information or positive social interactions (Oulasvirta et al., 2012). The receipt of these rewards can be unpredictable and the level of reward highly variable. This is likely to promote stronger associations, increasing the frequency and persistence of a behaviour being executed (Egel, 1980; Morford, Witts, Killingsworth, & Alavosius, 2014, Williams, 2006).

Checking behaviours on mobile devices tend to be supported by interactive notifications, by which mobile devices frequently attempt to gain the user's attention. Rather than just displaying information, these notifications commonly include a request for the user to select a function from a set of options, much like a dialog box. These types of notifications can become strong cues that instigate particular habitual responses to clear them. Notifications act as frequent calls for users to perform a consistent action and can act as gateway to further app use (Oulasvirta et al., 2012), making them optimal habitual response cues. Any design changes to these cues are likely to disrupt the habits that users form around an interface. Design changes are believed to force users out of automatic behaviour, removing their attention from the task, and instead focusing it back on the design of the interface (Raskin, 2000).

To designers, it may seem that the concept of user habituation has already been considered in the design of interfaces through popular design guidelines and heuristics (Nielsen, 1993; Shneiderman & Plaisant, 2010; Thimbleby, 1985). Design guidelines such as consistency, predictability, and standardisation of presentation (Shneiderman & Plaisant, 2010) allude to the idea that an ideal interface environment allows users to develop and maintain an interaction habit. Whilst these rules have proven invaluable to contemporary interface design, they alone do not give us a sense of the potential impact that habit development has on performance. Furthermore, there is no indication of the time taken to create these behaviour patterns or the extent to which performance is impacted if these patterns are disrupted. The current paper aims to explore these aspects of interface habit development in more detail.

### 3. Aims & Hypotheses

The current research contributes key insight on fundamental user behaviours by quantifying how the process of habit formation and disruption through design affect the speed and accuracy of interactions. Crucially we investigate this both within lab based and in the wild settings. We present two studies that focus on forced choice tasks, reflecting the characteristics of alert and other dialog selection screens seen in most user interfaces (e.g. "OK" vs "Cancel" in alerts, or "Reply" vs "Mark as read" in messaging applications). Study 1 examines how the process of formation and disruption of habits affect performance in a simple two-alternatives forced choice task in the lab under controlled conditions. Study 2 extends from Study 1 by investigating whether the effects replicate to a common real-world context in the form of a smartphone interactive dialog notification as the forced choice task.

The following hypotheses are considered in this study:

**Hypothesis 1a -** Response time will improve from baseline levels when participants begin to form an interface habit.

**Hypothesis 1b -** Response time will be significantly negatively impacted when habit formation is disrupted.

**Hypothesis 2a -** Response accuracy will improve from baseline levels when participants begin to form an interface habit.

**Hypothesis 2b -** Response accuracy will be significantly negatively impacted when habit formation is disrupted.

## 4. Study 1 – Lab

### 4.1 Method

**4.1.1 Participants**. Nineteen participants (12 men, 7 women) with a mean age of 24 years were recruited from a European University. The study was given exemption from ethical review as it did not meet any of the requirements for full review due to the low risk involved. Participants were invited to participate via email and social media. A €10 gift voucher was given to each participant as an honorarium for participation.

**4.1.2 Materials and Task.** Participants completed a forced choice task on a laptop and were asked to make a choice of whether the correct name for the object in the image was displayed on the right or left of the screen. Two images were used in the study from the Snodgrass and Vanderwart corpus (Snodgrass & Vanderwart, 1980), with the order of their appearance randomised. For each trial, participants were instructed to select the direction using the arrow keys (either "left" or "right") that corresponded to the location of the correct name. This type of task was used to ensure that participants did not have any pre-existing habits before taking part in the study, which may be the case if using existing dialog notification designs or common interface forced choice tasks.

Participants completed a total of 240 trials balanced across the three phases of the experiment (80 trials per phase).While as little as 20 trials are enough to accurately measure reaction time in a two choice task (Hultsch, MacDonald, & Dixon, 2002), 80 trials is usually required to achieve a reaction time plateau (Parkin, Kerr, & Hindmarch, 1997), indicating a degree of automaticity. The presentation of the two images was balanced so that each image appeared an equal number of times in each phase (i.e. 40 times per image) and that the labels appeared an equal number of times in the top left and right-hand side of the screen. The order of presentation of the images and appearance of the labels were randomised for each participant. An example of the stimuli is shown in Figure 1.

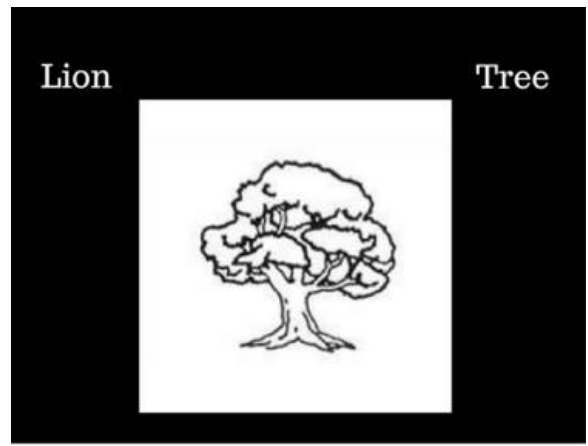

Figure 1. Example stimuli used in Study 1

**4.1.3 Experiment phases**. The experiment included three within-participant phases. In the *Baseline phase*, the corresponding label for each image appeared randomly in either the top left or right-hand corner of the screen. This condition identified a baseline response time for the task. In the *Habit phase*, the labels remained static and only the images were presented randomly (e.g. lion label was always on the left). In the *Disruption phase*, the side on which the label was displayed was randomised again, thus disrupting any learned response that the participants may have developed. Participants were not made aware of changes in phase, which happened in the background after 80 trials were completed.

This type of response time task is typical in tests of procedural memory (e.g. Knopman & Nissen, 1991), which have established that improvements in reaction time arise from an increase in automaticity. Making the placement of the labels consistent in the *Habit phase* facilitates users to map between the context (the image) and the response (the direction matching the label) in procedural memory, supporting repeated encoding of context- response patterns that are needed for automaticity to form (Wood & Rünger, 2016). Although there is debate about whether a habit should be defined as the behaviour or the impulse to perform a behaviour (see Gardner, 2015b), in the current task the users' impulse and actions were aligned, allowing both definitions to apply.

**4.1.4 Measures.** The most common way to measure habit is through self-report measures such as the Self-Report Habit Index (Verplanken & Orbell, 2003), or the Self-Report Behavioural Automaticity Index (Gardner, Abraham, Lally, & de Bruijn, 2012), which is a validated subscale of the SRHI focusing specifically on automaticity. Yet these types of measures can have downsides. Awareness decreases as automaticity develops (MacLeod & Dunbar, 1988), making subjective reflection on the automatic process potentially less reliable. Because of this, measures like response time and accuracy have been used as successful alternatives in gauging automaticity (Logan, 1979; Poldrack et al., 2005). We therefore measure how long people take to select the label (Response Time) and the accuracy of their response (Response Accuracy) in each of the trials.

**4.1.5 Procedure.** Participants were welcomed to the lab, given information on the purpose of the study and gave informed consent. They then received task instructions and were asked to complete all three phases of the experiment (*Baseline, Habit* and *Disruption*). The trials were displayed on a laptop using Psychopy version 1.8 (Peirce, 2007). After completing all phases, participants were debriefed about the purpose of the study and thanked for taking part in the research.

**4.2 Results**

Each participant completed 240 trials, 80 for each of the three phases, with 4560 entries in total.

**4.2.1 Response Time.** Using a-priori screening suggestions (Baayen & Milin, 2010), entries with impossible reaction times for a discrimination task (< 5ms & > 5s) were removed. This initial screening removed 22 trials (0.48%), leaving 4538 trials. To assess the effect of phases in the response time analysis, 245 trials where participants gave an incorrect answer were removed (5.4%; 4,293 remaining), following general convention for this type of analysis (Howell, 2013). Descriptive statistics for each experimental phase are shown in Table 1.

| Experiment Phases | Means | SD |
|---|---|---|
| Baseline | 1158 | 490 |
| Habit | 722 | 382 |
| Disruption | 1184 | 557 |

**Table 1: Mean response time (ms) per experimental phase.**

The response time data was analysed using a linear mixed- effects model (LME) using the lme4 package (Bates, Mächler, Bolker, & Walker, 2015) in R (R Core Team, 2014) version 3.5.1 Feather Spray. This analysis allowed us to assess the impact of the fixed effects (in this case the within participant variable: Experiment Phases) on the participant's response time, whilst controlling for individual differences in participant performance through random effects (for comprehensive review see Barr, Levy, Scheepers, & Tily, 2013). In accordance with Barr et al. (2013), we ran the maximal model with by-participant random slopes, simplifying the model where needed to achieve convergence. The LME model includes Experiment Phases as a fixed effect (3 levels - *Baseline, Habit & Disruption*) and by-participant random slopes (random effect). As both the hypotheses refer to comparisons between the *Habit* and the other conditions, the *Habit* condition is used as the intercept for all comparisons.

Upon looking at the distribution and the homoskedacity of residuals from the original analysis run on the untransformed data, the model appeared highly stressed at longer reaction times, suggesting a strong influence of outliers on the model. We therefore used an inverse

Making & Breaking Interface Habits12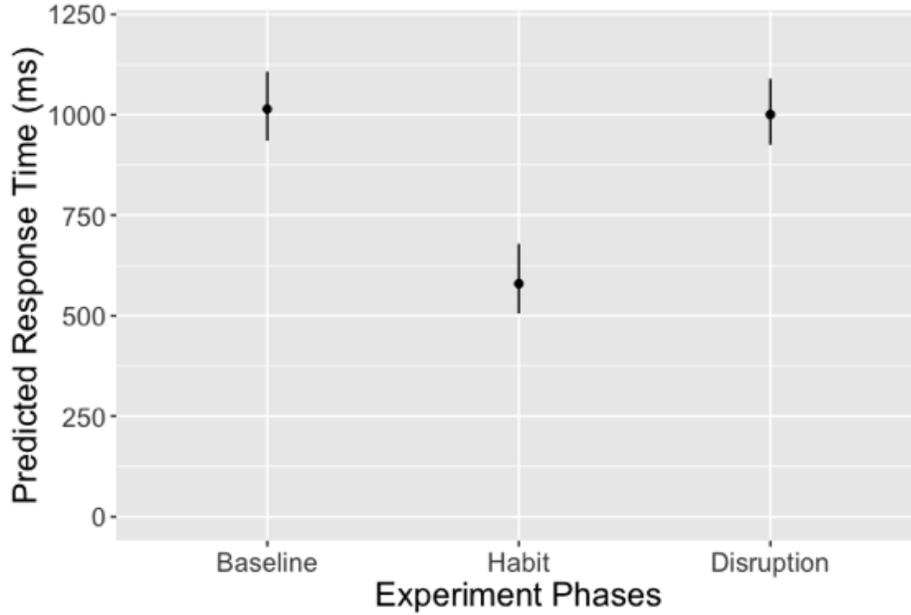

**Figure 2: Model estimates converted to predicted response time in milliseconds per phase. Line range shows 95% confidence interval. Data was transformed from inverse back into original scale for readability.**

transformation and model criticism of data outliers based on model residuals to limit the impact of outliers on the results (for in-depth discussion see Baayen & Milin, 2010). Using this method, 275 (6.41%) problematic trials were removed, leaving 4018 left for analysis. As suggested in (Baayen & Milin, 2010) the transformed reaction times were multiplied by negative 1000 to align estimates with the expected direction of interaction and to ensure the effect was visible at two decimal places.

| Predictors | Estimates | CI | t value | p |
|---|---|---|---|---|
| Habit (Intercept) | -1.72 | -1.97 – -1.47 | -13.60 | <0.001 |
| Baseline | 0.73 | 0.52 – 0.95 | 6.59 | <0.001 |
| Disruption | 0.72 | 0.50 – 0.94 | 6.38 | <0.001 |

**Table 2: Linear mixed effects regression analysis of the inverse of response times by experimental phase.**

Table 2 shows the results of the LME analysis after inverse transformations and negative multiplication. Figure 2, shows the predicted response times transformed back into the original scale to allow for greater readability and a more intuitive inspection of the differences between groups (produced by the Effects package version 4_0_3 Fox, 2003).

The analysis showed significantly longer response times in the *Baseline* phase when compared to the *Habit* phase (t = 6.48, p < .001), supporting Hypothesis 1a that habit formation leads to a significant improvement in speed compared to baseline levels. There was a similar statistically significant increase in the response times for the *Disruption* phase when compared to the *Habit* phase (t = 6.26, p < .001), supporting Hypothesis 1b that disrupting a habit leads to a decrease in speed.

**4.2.2 Response accuracy.** To assess the effect of condition on response accuracy, a logistic mixed-effects model was run. This analysis assesses the impact of the fixed effects (in this case Experiment Phases) on the log odds of selecting the correct item (incorrect: 0, correct: 1), whilst controlling for individual differences in performance through random effects. Descriptive statistics of response accuracy are included in Table 3.

| Phases | Means | SD |
|---|---|---|
| Baseline | .938 | .242 |
| Habit | .976 | .152 |
| Disruption | .924 | .265 |

Table 3: Mean proportion of accurate responses per experimental phase.

Like the LME model used in the response time analysis, the model includes Experiment Phases as a fixed effect (3 levels: *Baseline, Habit & Disruption*) and by participant random effects. Table 4 below shows the regression estimates transformed back into odd ratios for easier interpretation, while Figure 3 shows modelled probabilities (produced by the Effects package version 4_0_3 Fox, 2003) that a participant will get a correct answer in each phase.

| Predictors | Odd Ratios | CI | z value | p |
|---|---|---|---|---|
| Habit (Intercept) | 47.74 | 30.37 - 75.05 | 16.75 | <0.001 |
| Baseline | 0.51 | 0.28 - 0.95 | -2.14 | 0.032 |
| Disruption | 0.35 | 0.21 - 0.58 | -4.14 | <0.001 |

Table 4: Logistic mixed effects regression analysis of the odds of correct answer per experimental phase.

There was a significant decrease in correct answers during *Baseline* when compared to the *Habit* phase (z = -2.14, p < .05), supporting Hypothesis 2a that response accuracy is improved as habits form.

There was also a significant decrease in correct answers during *Disruption* when compared to the *Habit* phase (z = -4.14, p < .001), supporting Hypothesis 2b that disrupting a habit negatively impacts response accuracy.

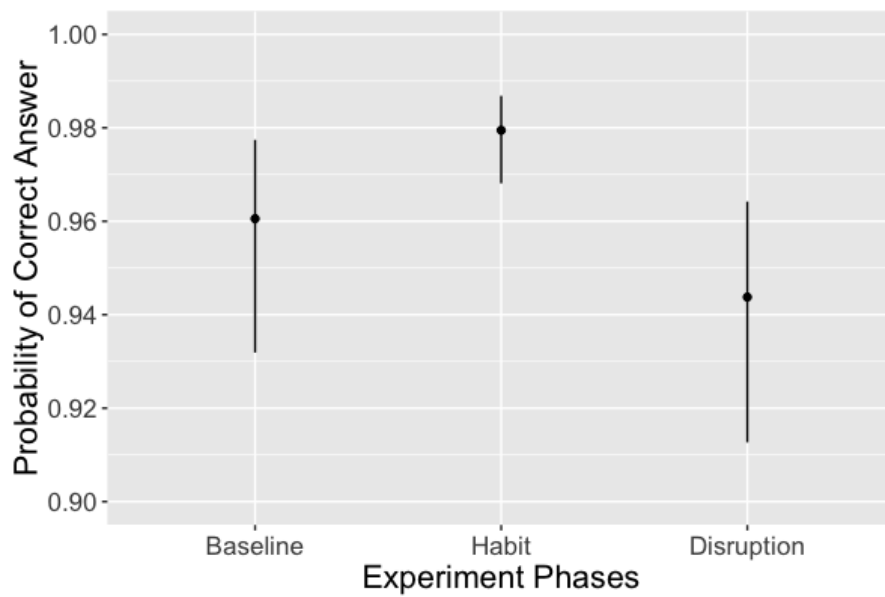

Figure 3: Probability of correct answers by phase. Line range shows 95% confidence interval.

### 4.3 Discussion

The experimental evidence of study 1 shows that, like other habits, allowing participants to form interface habits leads to significant gains in performance, as users became both more accurate and quicker at selecting the desired option. Importantly, we found that when habits becomes disrupted, any of the performance improvements are erased, returning to baseline levels. Overall the results highlight how moving context-response patterns to procedural memory is an effective way to improve performance and accuracy. When these patterns can no longer be used (i.e. habits are disrupted), performance gains are drastically reduced.

      The findings show strong effects of the interface habit formation process on performance that could be highly relevant to user interface behaviours. Yet these effects were elicited in a lab study, where users were interacting in a one-off session on a laptop-based interface. The task also lacked ecological validity, in that it did not resemble the everyday types of forced choice task encountered by users. The controlled lab setting is also not representative of the distraction-filled environment that is common when needing to make notification or dialog box choices. Study 2 therefore extends from Study 1 by running a similar study in the wild on smartphones, replacing Study 1's forced choice task with a more ecologically valid dialog notification task.

## 5. Study 2 – In the Wild

**5.1 Method**

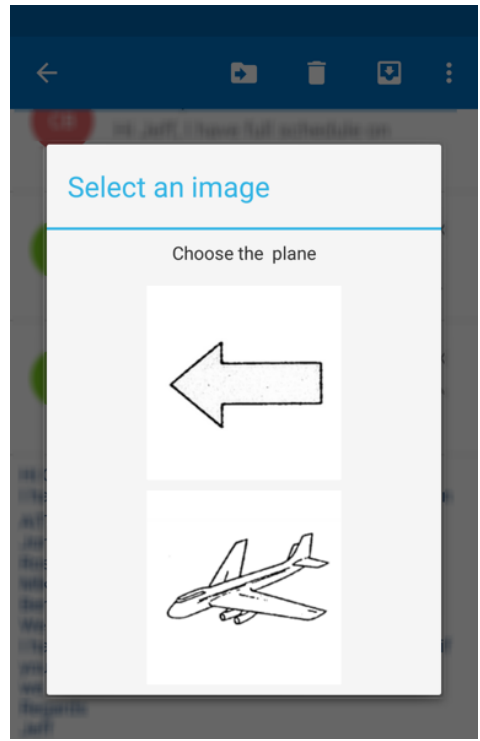

Figure 4: Example of notification dialog used in Study 2

**5.1.1 Participants.** 18 participants (8 men, 10 women) with a mean age of 29 years took part in the experiment as part of a wider study on user behaviour towards mobile notifications. The study was given full ethical approval before starting recruitment. Participants were recruited from two European Universities through email. They were given a £10 voucher as honorarium for their participation in the study. To be eligible to take part, participants needed to be over 18 years of age and own an Android device running version 4.0.3 or newer.

**5.1.2 Dialog task and materials.** The experiment application presented a series of notification-like dialogs to the user; an example is shown in Figure 4.

To ensure that any effect seen was not confounded by users' existing notification habits, we altered the design of the notifications used. Rather than using common text action buttons, participants were asked to select an image specified in the notification text from two images displayed. We also reversed the usual convention of two text action options placed horizontally to two image action options placed vertically. The 134 images used were taken from images by Snodgrass and Vanderwart (1980). The images appeared in a random order within a dialog that was placed on top of the existing device interface, obfuscating the user's current task. The dialog required the user to select the correct image before it could be cleared. The application was designed to only display the dialogs when the device detected that the user was currently actively engaging with the device. This was to ensure the user's attention is on the device at the time of the notification appearing.

**5.1.3 Experiment phases.** As in Study 1, the experiment included within participant experimental phases. In the *Baseline* phase all participants were presented with dialogs where the images that needed to be selected by user were randomly positioned. This acted as a baseline, assessing user performance before a habit could develop. In the *Habit* phase, each individual correct image was always placed in the same location (either in the top or bottom of the dialog) for each trial. The *Disruption* phase once again had the images appear randomly in either position. The phases were changed remotely via the server after at least 22 days to ensure users had the opportunity to complete enough interactions for each phase. Participants were not made aware of the phase change.

**5.1.4 Measures.** Similar to Study 1, Response Time and Response Accuracy were measured. Response Time was measured as the time difference between the logged time the dialogue was first presented to the logged time that a dialog was cleared. Response Accuracy was measured as whether the users selected the image they were asked to select in their initial response.

**5.1.5 Procedure.** At the start of the study, participants were directed to a website which contained participant information and a request for consent. Once consent was gained, participants were directed to instructions on how to download, install and request support for the application. To ensure that participants had sufficient opportunity to install the application and gain support if required, we allowed a 14-day interval between initial release of the recruitment call to the commencement of the first phase of the study.

Once a trial (i.e. a notification dialog) was completed and logged by the server the next trial would be generated and sent to the participant's phone. As mentioned above, each within-subject phase was administered remotely after at least 22 days via the server and deployed by the server to each device at the next available opportunity to ensure that all phases started on the same day. The duration of the phases was chosen to ensure each user completed enough interactions per phase. Participants completed an average of 130 interactions per phase, well above the 80 trials recommended to reach a response time plateau (Parkin et al., 1997) needed to signal acquired automaticity. At the end of the final (third) phase, a thank you message was pushed to participants, which included instructions on how to uninstall the applications from their personal devices. The study lasted for a total duration of 78 days.

**5.2 Results**

A total of 18 participants agreed to take part in the study, contributing a total of 6352 trials. We excluded participants who did not fully complete the study or who contributed too few trials to the *Baseline* condition (6 participants) leaving a total of 12 participants (5 men, 7 women) and 4712 trials. Impossible response times below 5ms and above 5s were also removed (565 trials), leaving an overall total of 4147 trials for analysis. Descriptive statistics are displayed in Table 5.

**5.2.1 Frequency of interactions.** On average, the remaining participants after removal interacted with a dialog notification between 2.9 and 7.6 times per day, with an overall mean of 5.0.

**5.2.2 Response time.** As with Study 1, incorrect answers were removed from the response time analysis, reducing the number of entries by 3.64% (151 entries). Descriptive statistics for response time are shown in Table 5.

| Phases | Means | SD |
|---|---|---|
| Baseline | 2488 | 744 |
| Habit | 2105 | 690 |
| Disruption | 2409 | 679 |

Table 5: Mean response time (ms) per experimental phase.

An LME model was used on the inverse of the data (for the same reasons as Study 1) and included Experiment Phase (3 levels: *Baseline, Habit* and *Disruption*) as a fixed effect with by participant random slopes as random effects. Evidence of model stress required further screening based on model criticism.

| Predictors | Estimates | CI | t value | p |
|---|---|---|---|---|
| Habit (Intercept) | -0.51 | -0.54 – -0.48 | -34.45 | <0.001 |
| Baseline | 0.08 | 0.06 – 0.10 | 7.90 | <0.001 |
| Disruption | 0.06 | 0.05 – 0.07 | 10.71 | <0.001 |

Table 6: Linear mixed effects regression analysis of the inverse of response times by phase.

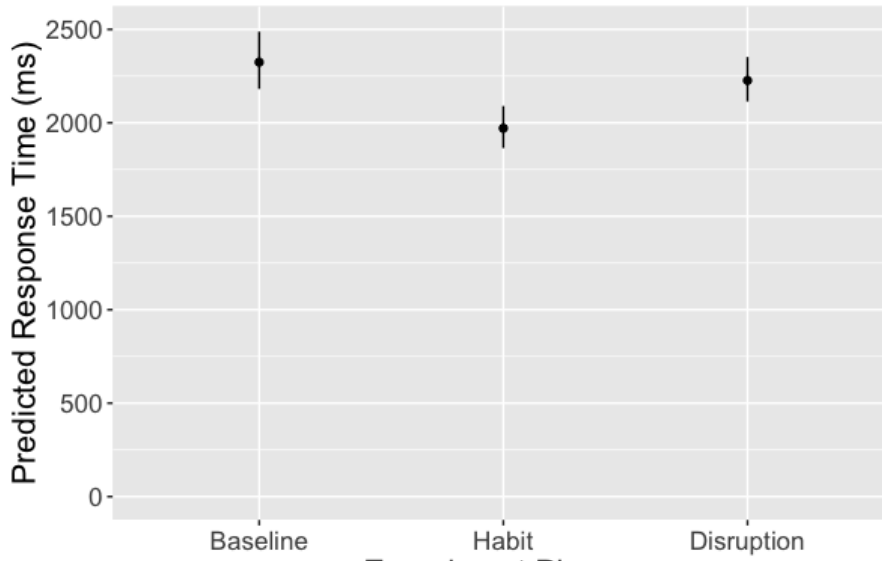

Figure 5: Modelled estimates converted to predicted response time in milliseconds per phase. Line range shows 95% confidence interval.

Sixty-seven (1.68%) entries were removed using this method, leaving 3929 trials. Table 6 shows the results of the LME analysis with the inverse transformation, while Figure 5 shows the modelled reaction times per phase untransformed.

Like in Study 1, the analysis showed significantly longer participant response times in the *Baseline* phase when compared to the *Habit* phase (t = 7.90, p < .001), again supporting Hypothesis 1a that habit formation leads to a significant improvement in speed compared to baseline levels. There was a similar statistically significant increase in the response times for the *Disruption* phase when compared to the *Habit* phase (t = 10.71, p < .001), again supporting Hypothesis 1b that disrupting a habit leads to an increase in response times.

| Phases | Means | SD |
|---|---|---|
| Baseline | .963 | .190 |
| Habit | .967 | .178 |
| Disruption | .960 | .195 |

**Table 7: Mean proportion of accurate responses by experimental phase.**

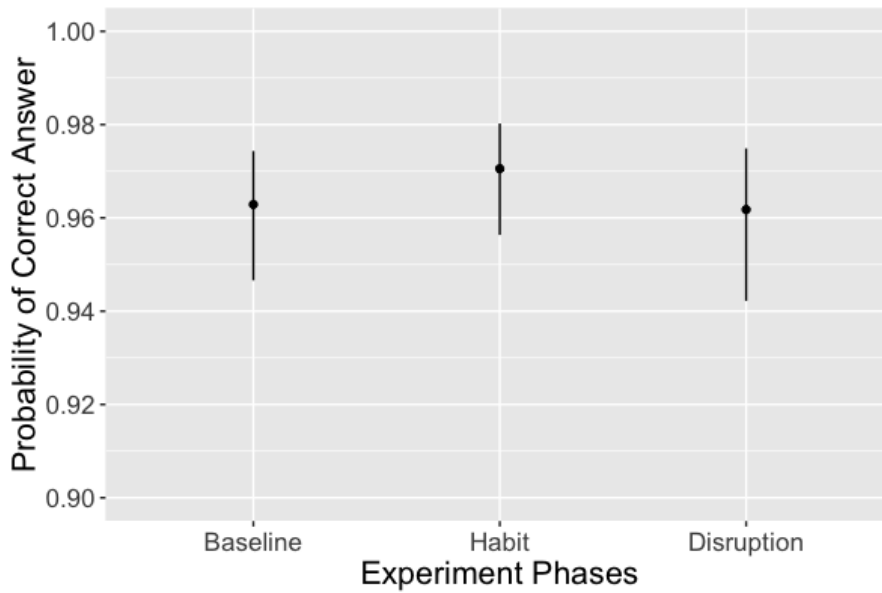

**Figure 6: Probability of correct answer per phase. Line range shows 95% confidence interval.**

**5.2.3 Response Accuracy.** Descriptive statistics for response accuracy are shown in Table 7.

Like in Study 1, a logistic mixed-effects model was run to assess the effect of phase on accuracy. This analysis again assesses the impact of the fixed effects (in this case Experiment Phases) on the log odds of selecting the correct item (incorrect: 0, correct: 1), whilst controlling for individual differences in performance through random effects. Table 8 shows the results of the logistic mixed effects analysis, while Figure 6 shows the probability of correct answer per phase. The model found no statistically significant difference between the *Habit* and either the *Baseline* (z = -0.72, p = .471) or *Disruption* phases (z = -1.40, p = .160), contradicting the findings in Study 1. Hypotheses 2a and 2b were therefore not supported in this study, as there was no improvement in the probability of getting a right answer in the *Habit* phase and subsequently no negative impact during the *Disruption* phase.

| Predictors | Odd Ratios | CI | z value | p |
|---|---|---|---|---|
| Habit (Intercept) | 31.82 | 21.79 – 46.46 | 17.91 | <0.001 |
| Baseline | 0.85 | 0.55 – 1.31 | -0.72 | 0.471 |
| Disruption | 0.76 | 0.52 – 1.11 | -1.40 | 0.160 |

Table 8: Logistic mixed effects regression analysis of the log odds of correct answer per phase.

### 5.3 Discussion

The results of Study 2 showed that, like Study 1, participants were significantly faster at correctly answering the dialogs when the interface allowed for the formation of habits. Similar to Study 1 this improvement was erased when the habit was disrupted, with participants returning to levels of performance similar to baseline performance. However, unlike Study 1 we found that accuracy did not improve when the interface changed compared to their baseline performance. These findings contradict previous research stating that habitual behaviour is less prone to errors than when a habit is not formed (Graybiel, 2008; Wood & Rünger, 2016).

Response times in Study 2 were longer than those in Study 1. This may be because of the in the wild nature of the study. Participants were interrupted with the dialog as opposed to expecting the forced choice task in a lab setting, which may have contributed to the response delay. Another reason may be that both studies varied in the visual design and device context. Study 1 was conducted on a laptop, allowing the labels and pictures in the study to be clearly visible to participants. Study 2 was conducted on a much smaller screen, with participants having to read more detailed written instructions on this display. This increased focus could have led to higher selection times compared to Study 1. Nevertheless, the clear difference in response time between the phases show how in the wild measures can still be useful in quantifying increased automaticity in response to a cue, providing a useful metric for interface habit development.

## 6. General Discussion

This work contributes insight into fundamental user behaviours, particularly on how creating an environment that allows for a habit to form can affect user task performance. The work focuses particularly on the process of formation and disruption of interface habits around option selection, such as those seen in dialog notifications on a number of devices (e.g. choosing "Reply" or "Archive" in an email application).

We show that an interface conducive to habit formation significantly improve the speed at which people select their desired option (Study 1 & 2). These findings echo those focused on other behaviours in the psychology literature which state that making a behaviour automatic can drastically improve performance (Aarts & Dijksterhuis, 2000; Dezfouli & Balleine, 2012; Logan, 1979). Importantly, we show that the creation of interface habits can not only be supported and disrupted in a single session in the lab, but also within the context of long-term everyday use. The process of interface habit formation occurred even when interactions were dispersed throughout a block of 22 days, at an average of 5 interactions per day.

Our findings show that the improvements resulting from the formation of interface habits are drastically reduced after the interface is changed as procedural memory is no longer involved. The lab study (Study 1) indicated that disrupting habits has significant detrimental effects on both speed and accuracy, while the in the wild study (Study 2) only saw a negative impact on speed. Once again, this supports previous psychological research on habits that describe the negative effects of habit disruption on speed (Anderson, Folk, Garrison, & Rogers, 2016; Lally, van Jaarsveld, Potts, & Wardle, 2010), with only partial support for effects in accuracy (Logan, 1979). The discrepancy in the findings for accuracy may have arisen due to limitations of a study conducted outside controlled conditions. The real-world mobile notification context may be a contributory factor in this. Dialogs were presented when performing other tasks, subsequently interrupting the user. This unexpected interruption may have thus led users to select the wrong image in error, thus inflating the number of errors made.

In both studies, the habit phase led to improvements in response time. This is believed to occur due to a shift of context- response associations into procedural memory (Wood & Neal, 2007). Once in procedural memory, participants no longer need to read the labels to allow them to select the correct action and can rely solely on their automatic reaction. While it could be argued that improvements in response time occurred because the habit phase was easier, the reality is that the task was only easier due to the increase in automaticity from the development of a habitual response. If a user was not able to form habits or move associations into procedural memory, they would not see such an improvement in performance. This deterioration or inability to learn new response patterns can be seen in patients suffering from damage to brain regions associated with procedural memory (Ackermann, Daum, Schugens, & Grodd, 1996) but is not seen when the damage is associated with other types of memory (Cohen & Squire, 1980).

Repeating the dialog behaviour using a consistent interface likely increases the strength of the stimulus-response associations within procedural memory. This learning increases the initiation and execution speed of the behaviour.

### 6.1 Implications

**6.1.1 The importance of supporting habits.** Our results highlight the performance benefits of supporting habit formation and maintenance in the design of interfaces. They also show the disruptive impact that breaking habits can have on users. Interface designers should look to maintain a consistency across designs for very common tasks to preserve gains in efficiency and effectiveness that users have developed when completing these tasks. This may not only increase the number of interactions with the application (Fogg & Hreha, 2010) but also make users less likely to change to a competitor application, as the cost of transition would be increased (Oulasvirta et al., 2012). The research also highlights how response times can be used to measure the degree of automaticity a user has when navigating an interface, which can be used to indicate the formation of interface habits.

**6.1.2 Adapting interfaces to account for user habits.** The current research highlights the power of response time measurements as a tool for gauging the interface habit formation of users. Our findings suggest that interface designers can use response times to measure the degree to which each individual user is forming an interface habit. With this type of granular data, different strategies can be employed to reduce the disruptive effects of necessary changes in the interface. For example, power users with strong interface habits may be presented with interim interfaces that can acclimatise users more gradually, reducing frustration while maintaining high performance levels. Furthermore, these measures could be used to ascertain when users are mindlessly interacting with important alerts, which would signify a need to change these interfaces so as to force users to pay closer attention.

**6.1.3 Habit hijacking as a dark pattern.** Although habit development benefits user performance, the nature of habits as automatic behaviours stimulated by particular cues can be exploited as a dark design pattern (Greenberg, Boring, Vermeulen, & Dostal, 2014). Pop-ups exemplify this unethical "Bait and Switch" behaviour by creating realistic looking windows and dialog boxes (cues) that are changed in their function to bring undesired results (such as opening malicious programs). This type of habit hijacking is also common in phishing scams, where malicious actors present a familiar interface (e.g. PayPal Website) in the hopes that users will provide important personal information (Dhamija, Tygar, & Hearst, 2006). There is a danger that unscrupulous designers may even change functionality dynamically, recording when is the best time to switch based on performance indicators of a habit being formed.

**6.1.4 Considerate disruption of habits.** Our work shows that disrupting habit formation can be significantly detrimental to user performance even with small changes in the design of notification dialog choices (e.g. their position). Although further work would need to be done to identify how our findings would scale to more general and major changes to interfaces (e.g. operating system interface changes), our work points to the need to carefully consider the effects of making such large-scale changes. Systems could be more aware of the need to supply extra confirmations or ways to undo functions for previously heavily habitual tasks identified on previous designs. This would make the new design more sympathetic to people's previous habitual actions.

**6.1.5 The importance of deliberate habit disruption.** It must be noted that there are also circumstances in which deliberate disruption of interface habits may be required. Cox, Gould,

Cecchinato, Iacovides, and Renfree (2016) describe such disruption as micro boundaries and suggest their use principally to encourage reflection and subsequent behaviour change. This type of deliberate disruption of built up habits should be considered where execution of habitual selection may lead to irreversible, and potentially dangerous outcomes (Norman, 1983). In safety critical circumstances such as medical surgery (Machno et al., 2015) or transport operation (Walker & Strathie, 2016), performing a habitual, often-practiced response in an interface without considering the consequences could be catastrophic. Disrupting these habits through micro boundaries may be necessary, even if it prevents users from progressing quickly.

**6.2 Limitations & Further Work**

Our work explores habit formation in the context of device interaction. In particular we focus on selection tasks in a lab-based computer interaction, following with an in the wild notification dialog experiment on smartphones. The differences in our experimental results between these two studies reflect the difficulty in transferring abstract theoretical concepts from the lab to a real-world setting, as the number of confounding variables increase. That said, our findings, especially those with reference to response time, were similar across the experimental contexts.

Throughout the studies we have used abstract forced choice tasks as a way to simulate user option selection. This was primarily for the purposes of study control, reducing the confounds that exist if we used existing interface selection tasks. The tasks chosen were also simple and only focused on two distinct choices. This was so as to support the habit forming process over the experiments as research has shown that overly convoluted tasks could interrupt this (Lally et al., 2010). We also altered the tasks over two studies, to increase ecological validity. Future work should investigate the role that complexity may play in the formation and disruption of interface habits as well as look to keep the tasks consistent across the studies to identify any impact this may have had on our study findings.

As the dialog conditions for study 2 could not be changed at the individual participant level, and due to some participants being delayed in setting up the application correctly on their device, participants were exposed to the baseline condition for variable amounts of time. To ensure this did not impact our findings, we removed participants that did not complete a satisfactory number of interactions during the baseline phase from the analysis. For the remaining participants, a similar number of interactions were collected across all three phases, ensuring a consistent set of responses for each condition in the analysis.

## 7. Conclusion

Our research examined the impact of creating and disrupting interface habits on interaction efficiency and effectiveness as measured by performance and accuracy. Our results indicate that people can develop option selection habits both in concentrated bouts of interaction in the lab, and during more dispersed dialog interactions in the wild. The development of a habit improved task performance speed and increased user accuracy, although accuracy benefits did not occur in the wild. We show that disrupting this habit development has serious consequences for user performance, leading people to be slower when selecting options. We also show that response times can be used effectively to differentiate between users who are and are not forming an interface habit, allowing for different actions to be targeted at each group. When developing interfaces, designers must be aware of this tendency to form habits in interaction and, through design, look to support habitual behaviours.

## 8. Acknowledgements

Making & Breaking Interface Habits 28This work was funded by the HW Wilson Foundation, the NUI Travelling Scholarship and Science Foundation Ireland ADAPT Centre 13/RC/2106).

## 9. Dataset & Analyses

The datasets and analyses used for the current research are available in the Open Science Framework repository at https://osf.io/w6n3e/.

Making & Breaking Interface Habits29# 10. References

Aarts, H., & Dijksterhuis, A. (2000). Habits as knowledge structures: Automaticity in goal-directed behavior. *Journal of Personality and Social Psychology*, *78*(1), 53–63. doi:10.1037/0022-3514.78.1.53

Ackermann, H., Daum, I., Schugens, M. M., & Grodd, W. (1996). Impaired procedural learning after damage to the left supplementary motor area (SMA). *Journal of Neurology, Neurosurgery, and Psychiatry*, *60*(1), 94–97.

Anderson, B. A., Folk, C. L., Garrison, R., & Rogers, L. (2016). Mechanisms of habitual approach: Failure to suppress irrelevant responses evoked by previously reward-associated stimuli. *Journal of Experimental Psychology. General*, *145*(6), 796–805. doi:10.1037/xge0000169

Baayen, R. H., & Milin, P. (2010). Analyzing reaction times. *International Journal of Psychological Research*.

Barr, D. J., Levy, R., Scheepers, C., & Tily, H. J. (2013). Random effects structure for confirmatory hypothesis testing: Keep it maximal. *Journal of Memory and Language*, *68*(3). doi:10.1016/j.jml.2012.11.001

Bates, D., Mächler, M., Bolker, B., & Walker, S. (2015). Fitting linear mixed-effects models using lme4. *Journal of Statistical Software*, *67*(1), 1–48. doi:10.18637/jss.v067.i01

Cohen, N. J., & Squire, L. R. (1980). Preserved learning and retention of pattern-analyzing skill in amnesia: dissociation of knowing how and knowing that. *Science*, *210*(4466), 207–210.

Cowan, B. R., Bowers, C. P., Beale, R., & Pinder, C. (2013). The Stroppy Kettle: An Intervention to Break Energy Consumption Habits. *In CHI'13 Extended Abstracts on Human Factors in Computing Systems*, 1485–1490.

Cox, A. L., Gould, S. J. J., Cecchinato, M. E., Iacovides, I., & Renfree, I. (2016). Design Frictions for Mindful Interactions: The Case for Microboundaries. In *Proceedings of the 2016 CHI Conference Extended Abstracts on Human Factors in Computing Systems - CHI EA'' '16* (pp. 1389–1397). New York, New York, USA: ACM Press. doi:10.1145/2851581.2892410

Dezfouli, A., & Balleine, B. W. (2012). Habits, action sequences and reinforcement learning. *The European Journal of Neuroscience*, *35*(7), 1036–1051. doi:10.1111/j.1460-9568.2012.08050.x

Dhamija, R., Tygar, J. D., & Hearst, M. (2006). Why phishing works. In *Proceedings of the SIGCHI conference on Human Factors in computing systems'' - CHI '06* (p. 581). New York, New York, USA: ACM Press. doi:10.1145/1124772.1124861

Elhai, J. D., Levine, J. C., Dvorak, R. D., & Hall, B. J. (2017). Non-social features of smartphone use are most related to depression, anxiety and problematic smartphone use. *Computers in Human Behavior*, *69*, 75–82. doi:10.1016/j.chb.2016.12.023

Fogg, B. J., & Hreha, J. (2010). Behavior Wizard: A Method for Matching Target Behaviors with Solutions. In T. Ploug, P. Hasle, & H. Oinas-Kukkonen (Eds.), *Persuasive Technology* (Vol. 6137, pp. 117–131). Berlin, Heidelberg: Springer Berlin Heidelberg. doi:10.1007/978-3-642-13226-1_13

Fox, J. (2003). Effect displays in R for generalised linear models. *Journal of Statistical Software*.

Gardner, B. (2015). Defining and measuring the habit impulse: response to commentaries. *Health Psychology Review*, *9*(3), 318–322. doi:10.1080/17437199.2015.1009844

Making & Breaking Interface Habits31Parkin, C., Kerr, J. S., & Hindmarch, I. (1997). The effects of practice on choice reaction time and critical flicker fusion threshold. *Human Psychopharmacology: Clinical and Experimental*, *12*(1), 65–70. doi:10.1002/(SICI)1099-1077(199701/02)12:1<65::AID-HUP838>3.0.CO;2-W

Peirce, J. W. (2007). PsychoPy--Psychophysics software in Python. *Journal of Neuroscience Methods*, *162*(1-2), 8–13. doi:10.1016/j.jneumeth.2006.11.017

Pinder, C., Vermeulen, J., Cowan, B. R., & Beale, R. (2018). Digital behaviour change interventions to break and form habits. *ACM Transactions on Computer-Human Interaction*, *25*(3), 1–66. doi:10.1145/3196830

Pinder, C., Vermeulen, J., Wicaksono, A., Beale, R., & Hendley, R. J. (2016). If this, then habit: Exploring context-aware implementation intentions on smartphones. In *Proceedings of the 18th International Conference on Human-Computer Interaction with Mobile Devices and Services Adjunct - MobileHCI'' '16* (pp. 690–697). New York, New York, USA: ACM Press. doi:10.1145/2957265.2961837

Poldrack, R. A., Sabb, F. W., Foerde, K., Tom, S. M., Asarnow, R. F., Bookheimer, S. Y., & Knowlton, B. J. (2005). The neural correlates of motor skill automaticity. *The Journal of Neuroscience*, *25*(22), 5356–5364. doi:10.1523/JNEUROSCI.3880-04.2005

R Core Team. (2014). *R: A language and environment forstatistical computing.* Vienna, Austria: R Foundation for Statistical Computing. Retrieved from https://www.r-project.org/

Raskin, J. (2000). *The humane interface: New directions for designing interactive systems*. Reading, Mass: Addison Wesley.

Renfree, I., Harrison, D., Marshall, P., Stawarz, K., & Cox, A. (2016). Don't Kick the Habit: The Role of Dependency in Habit Formation Apps. In *Proceedings of the 2016 CHI Conference Extended Abstracts on Human Factors in Computing Systems - CHI EA'' '16* (pp. 2932–2939). New York, New York, USA: ACM Press. doi:10.1145/2851581.2892495

Shneiderman, B., & Plaisant, C. (2010). *Designing the User Interface: Strategies for Effective Human-Computer Interaction* (5th ed.). Boston MA, United States: Pearson Education India.

Snodgrass, J. G., & Vanderwart, M. (1980). A standardized set of 260 pictures: norms for name agreement, image agreement, familiarity, and visual complexity. *Journal of Experimental Psychology. Human Learning and Memory*, *6*(2), 174–215. doi:10.1037/0278-7393.6.2.174

Stawarz, K., Cox, A. L., & Blandford, A. (2015). Beyond Self-Tracking and Reminders: Designing Smartphone Apps That Support Habit Formation. In *Proceedings of the 33rd Annual ACM Conference on Human Factors in Computing Systems - CHI'' '15* (pp. 2653–2662). New York, New York, USA: ACM Press. doi:10.1145/2702123.2702230

Stawarz, K., Rodriguez, M. D., Cox, A. L., & Blandford, A. (2016). Understanding the use of contextual cues: design implications for medication adherence technologies that support remembering. *DIGITAL HEALTH*.

Thimbleby, H. (1985). User interface design: generative user engineering principles. In *Fundamentals of human–computer interaction* (pp. 165–180). Elsevier. doi:10.1016/B978-0-12-504582-7.50019-6

Van Deursen, A. J. A. M., Bolle, C. L., Hegner, S. M., & Kommers, P. A. M. (2015). Modeling habitual and addictive smartphone behavior. *Computers in Human Behavior*, *45*, 411–420. doi:10.1016/j.chb.2014.12.039